\documentclass[aps,prl,twocolumn,superscriptaddress,groupedaddress]{revtex4}
\usepackage{graphicx,amsmath,amssymb,amsfonts,latexsym,color,dcolumn,bm,mathtools}
\usepackage{pbox}
\usepackage{booktabs}

\newcommand{\beq}{\begin{equation}}
\newcommand{\eeq}{\end{equation}}
\newcommand{\bea}{\begin{eqnarray}}
\newcommand{\eea}{\end{eqnarray}}

\providecommand{\bra}[1]{\langle #1 \rvert}
\providecommand{\ket}[1]{\lvert #1 \rangle}

\providecommand{\kket}[1]{\lvert\lvert #1 \rangle}

\providecommand{\so}[1]{#1}

\providecommand{\sub}[1]{_{\text{#1}}}

\graphicspath{ {/} }

\usepackage{tikz}
\usepgflibrary{arrows}
\usetikzlibrary{decorations}
\usetikzlibrary{decorations.pathmorphing,patterns}
\usetikzlibrary{scopes}
\usetikzlibrary{shapes, matrix}
\begin{document}

\title{Fano-Liouville Spectral Signatures in Open Quantum Systems} 
\author{Daniel Finkelstein-Shapiro}
\affiliation{Department of Chemistry and Biochemistry, Arizona State University, Tempe AZ 85282}
\affiliation{Sorbonne Universit\'es, UPMC Univ Paris 06, UMR 7616, Laboratoire de Chimie Th\'eorique, F-75005, Paris, France}
\affiliation{CNRS, UMR 7616, Laboratoire de Chimie Th\'eorique, F-75005, Paris, France}
\author{Ines Urdaneta}
\affiliation{Sorbonne Universit\'es, UPMC Univ Paris 06, UMR 7616, Laboratoire de Chimie Th\'eorique, F-75005, Paris, France}
\affiliation{CNRS, UMR 7616, Laboratoire de Chimie Th\'eorique, F-75005, Paris, France}
\affiliation{Institut des Sciences Mol\'eculaires d'Orsay, B\^atiment 350, UMR8214, CNRS-Universit\'e Paris-Sud, 91405 Orsay, France}
\author{Monica Calatayud}
\affiliation{Sorbonne Universit\'es, UPMC Univ Paris 06, UMR 7616, Laboratoire de Chimie Th\'eorique, F-75005, Paris, France}
\affiliation{CNRS, UMR 7616, Laboratoire de Chimie Th\'eorique, F-75005, Paris, France}
\affiliation{Institut Universitaire de France, France
}
\author{Osman Atabek}
\affiliation{Institut des Sciences Mol\'eculaires d'Orsay, B\^atiment 350, UMR8214, CNRS-Universit\'e Paris-Sud, 91405 Orsay, France}
\author{Vladimiro Mujica}
\affiliation{Department of Chemistry and Biochemistry, Arizona State University, Tempe AZ 85282}
\author{Arne Keller}
\affiliation{Institut des Sciences Mol\'eculaires d'Orsay, B\^atiment 350, UMR8214, CNRS-Universit\'e Paris-Sud, 91405 Orsay, France}

\begin{abstract}
The scattering amplitude from a set of discrete states coupled to a continuum became known as the Fano profile, characteristic for its asymmetric lineshape and originally investigated in the context of photoionization. The generality of the model, and the proliferation of engineered nanostructures with confined states gives immense success to the Fano lineshape, which is invoked whenever an asymmetric lineshape is encountered. However, many of these systems do not conform to the initial model worked out by Fano in that i) they are subject to dissipative processes and ii) the observables are not entirely analogous to the ones measured in the original photoionization experiments. In this letter, we work out the full optical response of a Fano model with dissipation. We find that the exact result for the excited population, Raman, Rayleigh and fluorescence emission is a modified Fano profile where the typical lineshape has an additional Lorentzian contribution. Expressions to extract model parameters from a set of relevant observables are given. 
\end{abstract}

\maketitle


In a set of seminal papers spanning from 1935 to 1961, Beutler \cite{Beutler1935}, Fano \cite{Fano1935,Fano1961} and Friederichs \cite{Friedrichs1948} laid the basis of the theory to describe the absorption lineshapes of atomic photoionization experiments. These lineshapes present marked asymmetries which could not be explained by a simple Lorentzian resonance. The explanation was attributed to an interference between two photo-ionizaton pathways: one where the atom is ionized directly from its ground state and one where it is first excited to a higher discrete state which then ionizes (auto-ionizing states). The minimal Fano model consists in a discrete excited state coupled to a continuum set of excited states, both types of states being reachable by photo-excitation from the ground state.     
The resulting photo-fragmentation cross-section as a function of the excitation laser frequency $\omega_L$ is known as the Beutler-Fano or Fano profile:
\begin{equation}
h(\epsilon;q)=\frac{(q+\epsilon)^2}{\epsilon^2+1},
\label{eq:Fano-classic}
\end{equation}
where $q$ is the ratio of the transition dipole moment of the ground-discrete and ground-continuous transitions, and $\epsilon=(\hbar\omega_{L}-E_e)/\gamma$ where $E_e$ is the energy of the discrete state relative to the ground state and $\gamma=n\pi V^2$ is the linewidth of the excited state, induced by its coupling (per unit of energy) $nV^2$ to the continuum set of states, $n$ being the density of states.

Since the original photoionization experiments, the Beutler-Fano profile has been observed in an ever increasing variety of physical systems, and in particular in  nanoscale structures \cite{Mirosh2010,Lucky2012}. These include plasmonic nanostructures \cite{Pakizeh2009,Lucky2010,Hsu2014}, quantum dots, decorated nanoparticles~\cite{Lombardi2010} and spin filters \cite{Song2003}, to name a few. Although the Fano theory was built on a scattering framework where the observable was the population on the continuum (i.e. the ionized electrons), the result continued to be applied (with remarkable success) to dissipative, non-scattering systems where the observable was not always the population in the continuum of states. As noted by A. E. Miroshnichenko in 2010, {\it ``a suitable theory for a quantitative description of these cases is still lacking''} \cite{Mirosh2010}. 

 Some  attempts to supplement the Fano model with  dissipation processes have already been made. First by Fano himself in 1963~\cite{Fano1963} in a less-known paper, where the objective was to model pressure lineshape broadening by atomic collisions. In this paper, only pure dephasing was considered which amounts to considering solely elastic collisions. In the eighties, Agarwal et al.~\cite{Agarwal1982} included population relaxation from the discrete excited state to an additional discrete state to model the competition between atomic photo-ionization and spontaneous emission. 
K.~Rzyzewski~ and J. H. Eberly~\cite{Eberly1983}  considered pure dephasing using a Wiener-Levy stochastic process to describe phase fluctuations. 
More recently Kroner et al.~\cite{Kroner2008} and Zhang et al.~\cite{Zhang2011} included a general dissipation process for a system of semiconductor quantum dots but the equations were solved approximatively by neglecting the population on the continuum set of states. It is worth noting that these three last works focus on intense field effects, which will not be adressed in the present letter and will be the subject of a future work.

In this letter, we solve exactly the Fano dissipation problem for weak field, fulfilling the conditions for absorption, Raman, Rayleigh and fluorescence spectroscopies. Within the wideband approximation to describe the coupling to the continuum, we obtain explicit and simple expressions for all optical observables, such that they can be used explicitly by experimentalists, notably to directly obtain the system parameters. For this, we solve the  dynamics of the quantum system with energy levels in a Fano like configuration and coupled to a bath which induces excited states relaxation and pure dephasing. The system density matrix evolution is described in the Born-Markov approximation, by the Liouville equation in Lindblad form~\cite{Lindblad1976}.

Although our formalism is general and can be applied to a variety of systems, an important motivation is the type of architectures encountered in light harvesting systems called Gr{\"a}tzel or dye-sensitized solar cells~\cite{OReagan1991,Hagfeldt2010}, where a molecule is adsorbed to a semiconductor surface. In most important physical realizations, the electronic ground state of this hybrid system is an isolated quantum state located in the semiconductor energy gap while the excited states can be considered as superpositions of localized molecular excited states and delocalized semiconductor conduction band states~\cite{Duncan2007,Hagfeldt2010}. Evidence suggests that such a model with minor modifications could also account for molecules on metal nanoparticles~\cite{Lombardi2008,Lombardi2009,Lombardi2010}. Be it for solar energy applications, electronics or sensors, there is strong evidence that the details of the interface distinguish functioning from non-functioning devices \cite{Galoppini2004}.

The energy levels of our model, along with the possible transitions, are shown in Figure 1. 
A discrete excited state $\ket{e}$ with energy $E_e$ is coupled to a continuum of states $\ket{k}$ with energy $E_k$. These states can be reached from a ground state $\ket{g}$ through laser excitation. 
A sub-manifold (typically vibrational) $\ket{\nu}$ with energy $E_\nu$  is included in the electronic ground state to open  inelastic scattering channels as  $\ket{\nu} \rightarrow \ket{\nu'}$.  
\begin{figure}[ht]
\centering
\begin{tikzpicture}[scale=0.7]
\draw[thick] (0,0) -- (4cm,0);
\draw  (0,0.25cm) -- (4cm,0.25cm) node[right]{$\ket{\nu}$};
\draw  (0,0.5cm) -- (4cm,0.5cm);
\draw[thick] (0,3cm)--(4cm,3cm) node[right]{$\ket{e}$};
\draw[fill=gray]  (6cm,1cm) rectangle (7cm,5cm) node[midway,right]{$\ket{k}$} ;
\draw[->,thick] (2cm,0) --(2cm,2.9cm) node[midway, right] {$\mu_{\nu e}$};
\draw [->,thick] (3cm,0.5)--(6.25cm,3.25cm)
node[midway, above] {$\mu_{\nu k}$};
\draw[<->,thick] (3cm,3.1cm) to[out=45,in=135] node [sloped, above] {$V(k)$} (6.5cm,3.1cm);
\draw [->,thick,decorate,decoration=snake] (6.25cm,2.75cm)--(3cm,0)
node[midway,sloped ,below,] {$\Gamma(k,\nu)$};
\draw [->,thick,decorate,decoration=snake] (0.25cm,0.5cm)--(0.25cm,0cm) node[left=0.25cm, midway ] {$\Gamma_{\text{vib}}$};
\node (myfirstpic) at (4.4cm,7.5cm) {\includegraphics[scale=0.101]{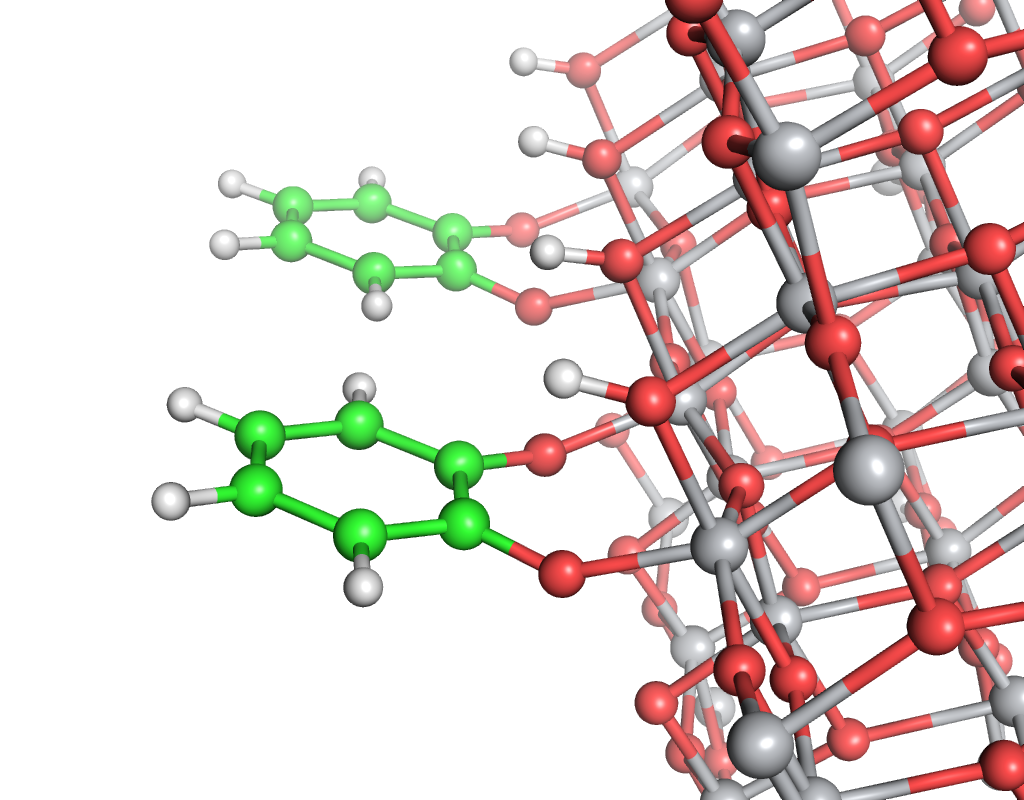}};
\end{tikzpicture}
\caption{\label{fig:FanoDissip} Energy levels and transitions of a Fano model with dissipation, including a vibrational ground state manifold, a discrete excited state and a continuum (bottom). A particular realization of a Fano system with dissipation is a molecule adsorbed on a metal oxide semiconductor, here two catechol molecules on a (101) anatase TiO$\sub{2}$ surface (top)}
\end{figure}
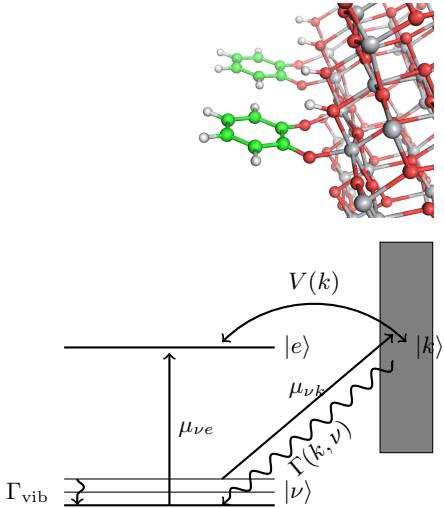
The Hamiltonian is:
\begin{align}
&H=H_0+H_V+H_F \\
&H_0=\sum_{\nu}E_{\nu}\ket{\nu}\bra{\nu}+E_e\ket{e}\bra{e}+\int dk E_k\ket{k}\bra{k} \nonumber \\
&H_V=\int dk \left[V(k)\ket{e}\bra{k}+V^*(k)\ket{k}\bra{e}\right] \nonumber \\
&H_F=F\sum_{\nu} \left[\mu_{\nu e}\cos(\omega_L t)\ket{\nu}\bra{e}+\mu_{\nu e}^*\cos(\omega_L t)\ket{e}\bra{\nu}\right] \nonumber \\
&+F\sum_{\nu} \int dk \left[\mu_{\nu k}\cos(\omega_L t)\ket{\nu}\bra{k}+\mu_{\nu k}^*\cos(\omega_L t)\ket{k}\bra{\nu}\right]
\label{eq:Hamiltonian}
\end{align}
where $H_0$ is the site Hamiltonian, $H_V$ is the coupling of the excited state to the continuum, for simplicity, in the following,  we will consider that $V(k) = \bra{e}H_V\ket{k}$ is real. $H_F$ is the interaction with the incident radiation field of frequency $\omega_L$, allowing transitions from the ground state to the discrete excited state $\nu \leftrightarrow e$ and to the continuum of states $\nu \leftrightarrow k$, $\mu_{ij}=\bra{i}\mu \ket{j}$ is the transition dipole moment between states $i$ and $j$ and $F$ is  the field amplitude. 

The originality of our model consists in taking into account a possible population relaxation from the continuum to the ground state at a rate $\Gamma(k,\nu)$, and pure dephasings.
When the photo-excitation involves an electron, this relaxation is mainly the result of the electron-hole Coulombic attraction followed by  the  thermal relaxation through the phonons of the environment. We phenomenologically capture this dissipation process using a  superoperator of Lindblad~\cite{Lindblad1976} form, which ensures the trace-preserving and complete positivity of the dynamical map, and solve the evolution of the density matrix with Liouville's equation:
\begin{equation}
\frac{\partial \rho}{\partial t}=\mathcal{L}(t)\rho
\label{eq:LiouvilleEq}
\end{equation}
where $\mathcal{L}(t)=\mathcal{L}_H(t)+\so{L}^D$, with $\hbar\mathcal{L}_H = -i(\openone \otimes H^T(t) - H(t)\otimes \openone)$, $L^D=\so{L}^D_{k}+\so{L}^D_{\text{vib}}+L^D_{\text{pure}}$ 
\begin{eqnarray}
&\so{L}^D_{k}=\sum_\nu\int dk \Gamma(k,\nu) \Big\{ A(k, \nu)\otimes A(k,\nu)  \nonumber \\
& - \frac{1}{2}\left[1\otimes A^{\dagger}(k, \nu)A(k,\nu) + A^{\dagger}(k, \nu)A(k,\nu)\otimes 1\right]\Big\},
\label{eq:dissipation_k}
\end{eqnarray}
\begin{eqnarray}
&\so{L}^D_{\text{vib}}=\sum_{\nu \neq 0} \Gamma_{\text{vib}} \Big\{A(\nu, 0)\otimes A(\nu,0) \nonumber \\
&- \frac{1}{2}\left[1\otimes A^{\dagger}(\nu, 0)A(\nu,0) + A^{\dagger}(\nu, 0)A(\nu,0)\otimes 1\right]\Big\} 
\label{eq:dissipation_vib}
\end{eqnarray}
and
\begin{eqnarray}
&L^D_{\text{pure}} =-\sum_{\nu}\gamma_{e\nu}\big[\ket{e}\bra{e}\otimes\ket{\nu}\bra{\nu}+\ket{\nu}\bra{\nu}\otimes\ket{e}\bra{e}\big] \nonumber\\
&-\sum_{\nu}\int dk\gamma_{k\nu}\big[\ket{k}\bra{k}\otimes\ket{\nu}\bra{\nu}+\ket{\nu}\bra{\nu}\otimes\ket{k}\bra{k}\big] \nonumber \\
&-\int dk\gamma_{ke}\big[\ket{k}\bra{k}\otimes\ket{e}\bra{e}+\ket{e}\bra{e}\otimes\ket{k}\bra{k}\big],
\label{eq:pure-dephasing}
\end{eqnarray}
where $H^T$ denotes the transpose of $H$, $A(i,j)=\ket{j}\bra{i}$ are the jump operators and $\Gamma(k,\nu)$ is the population relaxation rate from state $\ket{k}$ to $\ket{\nu}$. The superoperator $L^D_{\text{vib}}$ relaxes the ground states vibrational manifold and $\Gamma_{\nu 0}$ is the population relaxation rate  within the electronic ground state manifold. $\gamma_{ij}$ are the pure dephasing rates for the $ij$ coherences. 
We have used the isomorphism $L\tilde{\rho} R \rightarrow L\otimes R^{T} \rho$,  where $\rho$ is the  column form of the matrix $\tilde{\rho}$ through the correspondence: $\ket{l}\bra{m} \leftrightarrow \ket{l}\otimes \ket{m} \equiv\kket{lm}$~\cite{Havel2003}.

Non-radiative and radiative transitions from the electronic excited state $\ket{e}$ to the ground states $\ket{\nu}$ exist. They have been left out for simplicity, although their inclusion can be done straigthforwardly. Furthermore, in most cases of interest, the electronic coupling $V$ induces an effective population relaxation rate $2\pi n V^2/\hbar$ associated to electron injection into the semiconductor band which is much larger than the $e\rightarrow \nu$ rate corresponding to the direct relaxation mechanism, hence justifying this simplification~\cite{SM}. 

The optical response (absorption and emission spectra) is obtained through the Fourier transform of  the field two-times correlation function which in the far field is proportional to the dipole  two-times correlation function.  
Using the quantum regression theorem~\cite{Lax1963,Agarwal1974,CohenTannoudji2012},  the emitted light differential scattering cross-section can be written in the steady-state of the system as~\cite{Johansson2005,Xu2004}:
\begin{equation}
\begin{split}
&\frac{d^2\sigma}{d\Omega d(\hbar \omega)}= A(\theta)\times \\
&\times \sum_{a \in e,k}\sum_{b \in \nu} \mu_{ab}^2 \sum_{r\in \nu,e,k} Re[\rho_{ra}G_{ab,rb}\big(-i(\omega-\omega_L)\big)],
\end{split}
\label{eq:optical-emission-equation}
\end{equation}
with $A(\theta) = \frac{\omega^4\sin^2\theta}{I_{\text{in}}8\pi^3c^3\epsilon_0\hbar}$,
$d\Omega$ the element of solid angle, $I_{\text{in}}$  the incident laser intensity, $\theta$ the polar angle in spherical coordinates, $\rho$  the steady state density matrix and $G(z)=(z\openone-i\underline{\Omega}_L-\so{L})^{-1}$ the resolvent, which is the Laplace transform of the evolution superoperator, corresponding to  the time-independent Liouvillian $L=e^{i\underline{\Omega}_Lt}\mathcal{L}(t)e^{-i\underline{\Omega}_Lt}$ in the rotating-wave approximation (RWA), which consists in removing resonant oscillating prefactors and discarding non resonant terms. The $\underline{\Omega}_L$ matrix is a diagonal matrix with $\pm \omega_L$ for excited(ground)-ground(excited) coherences, and zero elsewhere. In summary, in order to obtain the optical emission we need to calculate the resolvent and the steady-state density matrix, which we do next. An explicit expression for $G(z)$ is obtained by separating $L$ into its diagonal and non-diagonal parts. While the diagonal contribution leads to a trivial calculation of its resolvent, the non-diagonal contribution is worked out through a Dyson equation to all orders in the field-free coupling and to second order in the field interaction. As for the steady-state density matrix $\rho$, it appears as the kernel of the  Liouvillian within the RWA, that is $ (-i\underline{\Omega}_L-L)\rho=0$, which can in turn also be expanded to second order in the field, with the help of the previously calculated resolvent $G(z)$~\cite{SM}.  

In principle, the above method allows to calculate the optical response for arbitrary couplings and density of states. Nevertheless, for many materials, the wideband  approximation, which considers $k$-independent couplings and a constant density of states, can faithfully reproduce experimental measurements. This was the same approximation which allowed a closed analytic form in the original Fano model \cite{Fano1961} and which has been used in all previous partial attempts to solve the Fano--dissipative problem~\cite{Fano1963, Agarwal1982,Eberly1983,Kroner2008, Zhang2011}. We introduce the notation $\Gamma_{c\nu}=\Gamma(k,\nu)$ and $\mu_{\nu c} = \mu_{\nu k}$ for the $k$-independent parameters.

A straightforward but tedious calculation~\cite{SM} yields all the terms needed to calculate the absorption spectrum and the emission differential cross-section (Eq \eqref{eq:optical-emission-equation}). The absorption and emission profiles, which carry the $\omega_L$ dependence, can be expressed in a compact and simple form in terms of the following function:
\begin{equation}
f(\epsilon,q,\eta,\alpha)=\alpha\frac{(q+\epsilon)^2}{\epsilon^2+1}+\eta\frac{q+1}{\epsilon^2+1}
\label{eq:profile}
\end{equation}
which is a linear combination of a Fano profile (first term) and a Lorentzian (second term); $q$ being the Fano asymmetry parameter and $\epsilon$ the normalized incident laser energy. $\alpha$ (which can only take the values $0$ and $1$) and $\eta$ are weighting coefficients for the Fano and Lorentzian components, respectively.  

The sum of excited state populations ($\rho_{ee}+\int dk \rho_{kk}$) can be written as:
\begin{equation}
N_{\text{excited}}=B^{\text{abs}}f(\epsilon,q,\eta,\alpha)
\label{eq:absorption}
\end{equation}
where $B^{\text{abs}}$ is a constant~\cite{SM}, $\epsilon$ and $q$ are the standard Fano parameters ($\epsilon=(E_e-\hbar\omega_{L})/\gamma$, $\gamma=n\pi V^2$, $q=\mu_{0e}/(n\pi V \mu_{0c})$), $\alpha=1$, and $\eta=(\sum_{\nu}\Gamma_{c\nu})/(4\pi n V^2)$. We can obtain the extinction coefficient $\beta$ measured in optical absorption measurements using arguments of detailed balance~\cite{SM}:
\begin{equation}
\begin{split}
\beta&= \frac{n\pi \mu_{0c}^2}{c\epsilon_0\hbar}\hbar \omega_L  f(\epsilon,q,\eta=0,\alpha=1)
\end{split}
\label{extinction-coefficient}
\end{equation}
where $c$ is the speed of light and $\epsilon_0$ the vacuum permittivity.
The differential cross sections for the three processes Rayleigh, Raman and fluorescence can all be expressed referring to the same functional form:
\begin{equation}
\frac{d^2\sigma^i}{d\Omega d(\hbar \omega)}=A(\theta)B_{\nu}^i R(\omega,\omega_0^i,\Delta^i)f(\epsilon^i,q^i,\eta^i,\alpha^i)
\label{eq:emission-cross-section}
\end{equation}
where $i=$ Rayleigh, Raman or fluorescence. The expression for emission has 3 parts: $A(\theta)B_{\nu}^i$ is a prefactor~\cite{SM}, $R(\omega,\omega_0,\Delta)=\frac{\Delta}{\pi}[(\omega-\omega_0)^2+(\Delta)^2]^{-1}$ is a normalized Lorentzian of central frequency $\omega_0$ and half-width $\Delta$ which gives the emission lineshape (for a fixed incident laser frequency), and $f(\epsilon,q,\eta,\alpha)$ which gives the profile - the signal integrated over the emitted frequency - and carries all the dependence of the incident laser frequency $\omega_L$. 

\begin{figure}[h]
	\includegraphics[width=0.5\textwidth]{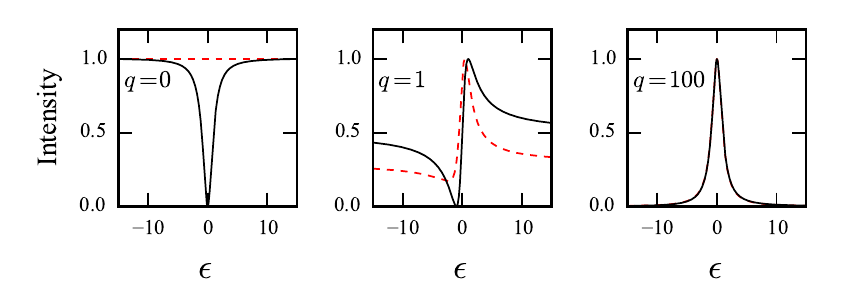}
	\caption{Characteristic function for absorption or Raman emission for $\eta=0$ (solid black) and $\eta=1$ (red dashed) for $q=0,1,100$ (intensities are normalized).}
\label{fig:Fano}
\end{figure}

We first discuss the characteristic function $f$ of the population given by equation~\ref{eq:absorption}. The result is the combination of a standard Fano profile, and a Lorentzian. The signature of the relaxation process is embodied in the parameter $\eta$. Its physical meaning is the ratio of the relaxation rate to the injection rate into the continuum. When $\eta \gg 1$ the relaxation quenches the population of the continuum and therefore suppresses the Fano interference signal giving rise to a pure Lorentzian lineshape. When $\eta \ll 1$, we recover the original Fano profile. Figure \ref{fig:Fano} shows normalized population profiles as given by Eq.~\eqref{eq:absorption}, for several values of $q$ and $\eta$. As $\eta \rightarrow 0$, the curve approaches the standard Fano profile while $\eta=1$ shows the Fano with a Lorentzian contribution for different values of the $q$ parameter. 

For the emission processes, we separate the full optical response given by Eq.~\eqref{eq:optical-emission-equation} into several components according to their central emission frequency $\omega_0^i$ of the lineshape functions $R(\omega,\omega_0^i,\Delta^i)$. The Rayleigh process corresponds to $\omega_0^{\text{Ray}}=\omega_L$ and the Raman process to $\omega_0^{\text{Ram}}=\omega_L-\omega_{\text{vib}}$, ($\hbar\omega_{\text{vib}}=E_{1}-E_{0}$). Contrary to the Rayleigh and Raman scattering which are coherent processes, fluorescence corresponding to a radiative transition from excited population is incoherent. This stationary population can only exist if in addition to population relaxation, pure dephasing processes are taken into account ($\gamma_{ij}\neq 0$ in Eq.~\eqref{eq:pure-dephasing}). The population of the discrete excited state leads to an emission at $E_e/\hbar$ and at $E_e/\hbar-\omega_{\textbf{vib}}$ central frequencies. The continuum states are populated at the laser frequency $\omega_L$ and therefore give a fluorescence emission with central frequencies at $\omega_L$ and $\omega_L-\omega_{\textbf{vib}}$. We note that the fluorescence from the continuum states occurs at the same central frequencies as the Rayleigh and Raman scattering, but can be distinguished by their respective lineshape widths $\Delta^{i}$. For Rayleigh, the width comes from the laser linewidth, $\Delta^{\text{Ray}}=\delta$, and for the Raman the width is given by the inverse lifetime of the vibrationally excited state $\nu=1$, $\Delta^{\text{Ram}}=\Gamma_{\text{vib}}/2$. The width of the fluorescence lineshape is dominated by the excited state population lifetime: for the emission at $\omega_L$ the width is $\Delta^{\text{fluor},c0}=\sum_{\nu}\Gamma_{c\nu}$ and for the emission at $\omega_L-\omega_{\textbf{vib}}$ the width is $\Delta^{\text{fluor},c1}=\sum_{\nu}\Gamma_{c\nu}+\Gamma_{\text{vib}}/2$. All the parameters for each emission component given by Eq.~\eqref{eq:emission-cross-section} are collected in Table~\ref{table:parameters}.

\begin{table*}
\begin{tabular}{ | c | c | c | c c | c c |}\hline  	&	 \multicolumn{2}{c|}{Lineshape} 		&	 \multicolumn{4}{c|}{Profile}	  \\ \hline
 Model: 	&	 \pbox{20cm}{Center \\ frequency} 	&	Width	&	 \pbox{20cm}{Laser \\ frequency} 	&	 \pbox{20cm}{Asymmetry \\ parameter} 	&	 \pbox{20cm}{Fano \\ weight $\alpha$} 	&	 \pbox{20cm}{Lorentzian \\ weight $\eta$} \\[10pt] \hline
 Standard: 	&		&		&	 $\epsilon$ 	&	 $q$ 	&	 $1$ 	&	 $-$  \\ \hline 
Excited state: 	&		&		&	 $\epsilon$ 	&	 $q$ 	&	 $1$ 	&	 $\eta$  \\ 
populations 	&		&		&	 	&	  	&	  	&	   \\ \hline 
Rayleigh: 	&	$\omega_L$ 	&	$\delta$	&	 $\epsilon$ 	&	 $q$ 	&	 $1$ 	&	 $\frac{\mu_{0 e}^2}{n\mu_{0 c}^2}\eta$  \\ 
Raman: 	&	$\omega_L-\omega_{\text{vib}}$ 	&	$\Gamma_{\text{vib}}/2$	&	 $\epsilon$ 	&	 $q$ 	&	 $1$ 	&	 $\frac{\mu_{1 e}^2}{n\mu_{1 c}^2}\eta$  \\ \hline 
Fluorescence: 	&	$(E_e-E_0)/\hbar$	&	$n\pi V^2/\hbar+\gamma_{e0}$ 	&	 $\frac{\epsilon (n\pi V^2)}{n\pi V^2+\hbar\gamma_{e0}}$ 	&	 $\frac{q(n\pi V^2)}{n\pi V^2+\hbar\gamma_{e0}}$ 	&	 $0$ 	&	 $(\frac{n\pi V^2}{n\pi V^2+\hbar\gamma_{e0}})^2  \eta$  \\ 
 discrete	&	 $(E_e-E_1)/\hbar$	&	$n\pi V^2/\hbar+\gamma_{e1}+\Gamma_{\text{vib}}/2$	&	 $\frac{\epsilon (n\pi V^2)}{n\pi V^2+\hbar\gamma_{e0}}$ 	&	 $\frac{q(n\pi V^2)}{n\pi V^2+\hbar\gamma_{e0}}$ 	&	 $0$ 	&	 $(\frac{n\pi V^2}{n\pi V^2+\hbar\gamma_{e0}})^2   \eta$  \\ \hline 
Fluorescence : 	&	$\omega_L$ 	&	$\sum_{\nu}\Gamma_{c\nu}+2\gamma_{e0}$	&	 $\frac{\epsilon (n\pi V^2)}{n\pi V^2+\hbar\gamma_{e0}}$ 	&	 $\frac{q(n\pi V^2)}{n\pi V^2+\hbar\gamma_{e0}}$ 	&	 $1$ 	&	 $\frac{\gamma_{e0}^2}{(n\pi V^2/\hbar+\gamma_{e0})^2}\frac{1}{q^2+1}$ \\
continuum	&	 $\omega_L- \omega_{\text{vib}}$	&	 $\sum_{\nu}\Gamma_{c\nu}+\gamma_{e0}+\gamma_{e1}+\Gamma_{\text{vib}}/2$	&	 $\frac{\epsilon (n\pi V^2)}{n\pi V^2+\hbar\gamma_{e0}}$ 	&	 $\frac{q(n\pi V^2)}{n\pi V^2+\hbar\gamma_{e0}}$ 	&	 $1$ 	&	 $\frac{\gamma_{e0}^2}{(n\pi V^2/\hbar+\gamma_{e0})^2}\frac{1}{q^2+1}$ \\	\hline
\end{tabular}
\caption{Parameters for all optical processes depicted in Eqs.~\eqref{eq:absorption}~\eqref{eq:emission-cross-section}. The parameters are expressed as a function of $\epsilon=(\hbar \omega_L-E_e)/(n\pi V^2)$, $q=\mu_{0e}/(n\pi V \mu_{0c})$ and $\eta=\sum_{\nu}\Gamma_{c\nu}/(4n\pi V^2)$.}
\label{table:parameters}
\end{table*} 

The parameters of the model can be extracted from spectroscopic measurements. By fitting the profiles, we can get $\gamma$, $q$, $B^{\text{abs}}$, $B^{\text{Ram}}$, $B^{\text{Ray}}$, $F$ (experimental parameter), $\eta^{\text{Ray}}$, and $\eta^{\text{Ram}}$ (Table~\ref{table:parameters} and S.M.~\cite{SM}). From these, there are various ways to obtain the model parameters, for instance:
\begin{equation}
\begin{split}
\sum_{\nu=0}^1\Gamma_{c\nu}&=\frac{B^{\text{Ray}}}{(B^{\text{abs}})^2}\frac{F^2}{2\hbar},\; \sqrt{n}V=\sqrt{\frac{\gamma}{\pi}} \\
\sqrt{n}\mu_{0c}&=\sqrt{ \frac{B^{\text{Ray}}}{\pi B^{\text{abs}}}}  ,\; \mu_{0e}=\sqrt{\frac{B^{\text{Ray}}\gamma}{B^{\text{abs}}}}q   \\
\sqrt{n} \mu_{1c}&=\sqrt{\frac{B^{\text{Ram}}}{\pi B^{\text{abs}}}} ,\; \mu_{1e}= \frac{ 8 \eta^{\text{Ram}} \gamma B^{\text{abs}}B^{\text{Ram}} \hbar}{\pi B^{\text{Ray}} F^2}
\end{split}
\end{equation}

\begin{figure}
	\includegraphics[width=0.5\textwidth]{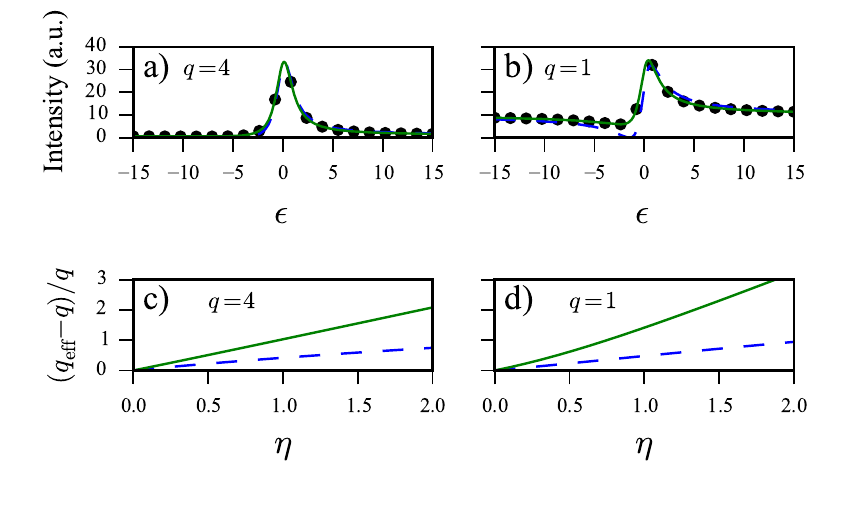}
	\caption{Comparison between different Fano models. Top: fits of the profile presented in this paper Eq.~\eqref{eq:profile} with a) $q=4$ and b) $q=1$, with $\eta=1$ by a standard (blue dashed) or a shifted Fano profiles (solid green). Bottom: relative differences between the extracted $q\sub{eff}$ and the actual $q$ value for c) $q=4$ and d) $q=1$ for the standard (blue dashed) or the shifted Fano profiles (solid green).}
\label{fig:Models}
\end{figure}
 
We now discuss how important is to include the Lorentzian term arising from dissipation in addition to the standard Fano profile. In Fig.~\ref{fig:Models}, we present a profile with our model Eq.~\eqref{eq:absorption} with a given set of parameters and fit it both with a standard Fano model $h(\epsilon;q_{\text{eff}})$ (Eq.~\eqref{eq:Fano-classic}) and with a shifted Fano model  $\frac{1}{N}[h(\epsilon;q_{\text{eff}})+D]$. The shifted Fano fitting is usually performed in optical measurements and corresponds to a constant background substraction. Below, we calculate the relative error between the extracted value of the Fano asymmetry parameter, which we call $q_{\text{eff}}$, and the original $q$ as a function of the Lorentzian weight factor $\eta$. We show that even when the fitting is very good, the extracted parameters can be off by a factor of two for $\eta=1$. This definitely shows the importance of including properly the dissipation in the Fano model. 

In conclusion, we expect that the equations will motivate the experimentalists to extract system parameters which were not accessible previously, from routine optical spectroscopies, characterizing with increasing precision the discrete-continuum interface both relevant in devices and interesting from a fundamental standpoint. Furthermore, our model serves as a stepping stone for further theoretical studies: optical response beyond the wide-band approximation (near band edges for example), strong field effects relevant for plasmonics, generalization to time-dependent laser pulse sequences relevant to non-linear 2D spectroscopy and application to real systems with the help of computational tools in order to obtain ab-initio the parameters of the model. 

\textbf{Acknowledgments} We thank Prof. J.H. Eberly for an overview of relevant literature. D.F.S. acknowledges the Research in Paris program for a fellowship. This work was financially supported by project NSF-ANR (ANR-11-NS04-0001 FRAMOLSENT program, NSFCHE-112489). This work was performed using HPC resources from GENCI- CINES/IDRIS (Grant 2015- x2015082131, 2014- x2014082131) and the CCRE-DSI of Universit\'e P. M. Curie. 

\bibliography{Fano}

\end{document}